\documentclass[english,showpacs,floatfix,preprint,10pt]{revtex4}
\usepackage[dvips]{graphicx}
\usepackage{longtable}
\usepackage{amsmath,amssymb}
\usepackage{dcolumn}
\usepackage{latexsym}
\usepackage{babel}
\usepackage[latin1]{inputenc}
\usepackage{color}
\usepackage[colorlinks]{hyperref}
\begin{document}

\title{Stability of the curvature perturbation in dark sectors'
mutual interacting models}
\author{Jian-Hua He$^{1}$, Bin Wang$^{1}%
$\footnote{wangb@fudan.edu.cn }, Elcio
Abdalla$^{2}$\footnote{eabdalla@fma.if.usp.br }} \affiliation{$^{1}$
Department of Physics, Fudan University, 200433 Shanghai, China}
\affiliation{$^{2}$ Instituto de F\'{\i}sica, Universidade de S\~ao
Paulo, CP 66318, 05315-970, S\~ao Paulo, Brazil.}

\begin{abstract}
We consider perturbations in a cosmological model with a small
coupling between dark energy and dark matter. We prove that the
stability of the curvature perturbation depends on the type of
coupling between dark sectors. When the dark energy is of
quintessence type, if the coupling is proportional to the dark
matter energy density, it will drive the instability in the
curvature perturbations; however if the coupling is proportional
to the energy density of dark energy, there is room for the
stability in the curvature perturbations. When the dark energy is
of phantom type, the perturbations are always stable, no matter
whether the coupling is proportional to the one or the other
energy density.
\end{abstract}


\maketitle

We are convinced by the fact that our universe is undergoing an
accelerated expansion driven by the so called dark energy (DE).
The leading interpretation of such a DE is the cosmological
constant with equation of state (EoS) $w=-1$.  Although the
cosmological constant is consistent with the observational data,
it presents apparently unsurmountable problems from the
theoretical point of view, which not only requires a severe fine
tuning of 120 digits to attain the actual value of the
cosmological constant, but also leads to the coincidence problem,
namely why the DE and the Dark Matter (DM) are comparable in size
exactly today\cite{REVIEWS_I}.

Dark Energy contributes a significant fraction of the content of
the universe. It is thus natural to consider its interaction with
the remaining fields of the Standard Model in the framework of
standard field theory. The possibility that DE and DM can interact
has been widely discussed recently
\cite{Amendola2000}-\cite{maartens}. It has been shown that
certain types of coupling between DE and DM can lead to a late
time attractor solution for the ratio of DM and DE densities
\cite{maartens1} and provide a mechanism to alleviate the
coincidence problem \cite{Amendola2000,Campo2006}. It has been
argued that an appropriate interaction between DE and DM can
influence the perturbation dynamics and affect the lowest
multipoles of the CMB spectrum \cite{Wang2007,Zimdahl2005}.
Arguments using galaxies structure formation suggested that the
strength of the coupling could be as large as the QED fine
structure constant \cite{Wang2007,Abdalla2007}. More recently, it
was shown that such an interaction could be inferred from the
expansion history of the universe, as manifested in, e.g., the
supernova data together with CMB and large-scale structure
information\cite{He08014233,feng,Guo2007}. In addition, it was
suggested that the dynamical equilibrium of collapsed structures
can be affected by the coupling between DE and DM
\cite{Bertolami2007}. The basic idea is that the virial theorem is
distorted by the non-conservation of mass caused by the
coupling\cite{Abdalla07101198}. Thermodynamical attempts to
understand the interaction between DE and DM has also been
proposed \cite{Pavon07120565}.

Recently there has been some concern about the stability of the
perturbations under DE and DM interaction \cite{maartens}, which
could represent a sharp blade in the heart of such interacting
models. In the original analysis the authors considered that the
energy exchange between DE and DM is proportional to the energy
densities of DM and total dark sectors. For the constant DE EOS
$w>-1$, it was found that the instability arises regardless of how
weak the coupling is. In this work we are going to reexamine the
stability of the curvature perturbation when dark sectors are
mutually interacting. We will concentrate on the interaction
between dark sectors in a linear combination of energy densities
of DE and DM, which is a more general phenomenological form in
describing the interaction \cite{waga,He08014233}. We will
restrict our investigation to constant EOS including $w>-1$ and
$w<-1$ cases.

We consider a two-component system with each energy-momentum tensor
satisfying 
\begin{equation}
\nabla_{\mu}T^{\mu\nu}_{(\lambda)}=Q^{\nu}_{(\lambda)}\label{tensor}
\end{equation}
where $Q^{\nu}_{(\lambda)}$ denotes the interaction between
different components and $\lambda$ denotes either the DM or the DE sector.
This equation can be projected on the time or on the space
direction of the comoving observer. Using the four velocity
$V_{\nu}$, it can be contracted into
\begin{equation}
V_{\nu}\nabla_{\mu}T^{\mu\nu}_{(\lambda)}= -\dot{\rho_{\lambda}} -
\theta (\rho_{\lambda} +
p_{\lambda})=V_{\nu}Q^{\nu}_{(\lambda)},\label{background1}
\end{equation}
which is the projection on the time direction of the comoving
observer. Above,
$\dot{\rho_{\lambda}}=V_{\nu}\nabla^{\nu}\rho_{\lambda}$ and $\theta
= \nabla^{\nu}V_{\nu}$ is the volume expansion rate. In order to
get the projection along the space direction, we can use
$h^{\tau}_{\nu}=\delta^{\tau}_{\nu}+V^{\tau}V_{\nu}$ on (\ref{tensor}) and
take the contraction
\begin{equation}
h^{\tau}_{\nu}\nabla_{\mu}T^{\mu\nu}_{(\lambda)}= (\rho_{\lambda}+
p_{\lambda})A^{\tau} +^{(3)}\nabla^{\tau}p_{\lambda}
=h^{\tau}_{\nu}Q^{\nu}_{(\lambda)}.\label{background2}
\end{equation}
where $A^{\tau}=V^{\mu}\nabla_{\mu}V^{\tau}$ is the acceleration. For
the homogeneous and isotropic universe, it requires
$^{(3)}\nabla^{\tau}p_{\lambda}=0$. Besides, for the DM particle,
its world line is the geodesic, $A^{\tau}=0$. Thus the spacial
part of $Q^{\nu}_{(\lambda)}$ vanishes, which means that in the
background there is no momentum transfer between dark sectors
\cite{maartens}. For the whole system the energy momentum
conservation still holds, satisfying $\Sigma_{\lambda}
\nabla_{\mu}T^{\mu\nu}_{(\lambda)}=0$, thus requiring
$Q^{0}_{DE}=-Q^{0}_{DM}$.

We choose the perturbed space-time
\begin{equation}
ds^2 = a^2[-(1+2\psi)d\tau^2+2\partial_iBd\tau
dx^i+(1+2\phi)\delta_{ij}dx^idx^j+D_{ij}Edx^idx^j],\label{perturbedspacetime}
\end{equation}
where
\begin{equation}
D_{ij}=(\partial_i\partial_j-\frac{1}{3}\delta_{ij}\nabla^2).
\end{equation}
The perturbed energy-momentum tensor reads
\begin{eqnarray}
\delta \nabla_{\mu}T^{\mu0}_{(\lambda)} &=&
\frac{1}{a^2}\{-2[\rho_{\lambda}'+3\mathcal
{H}(p_{\lambda}+\rho_{\lambda})]\psi+\delta
\rho_{\lambda}'+(p_{\lambda}+\rho_{\lambda})\theta_{\lambda}+3\mathcal
{H}(\delta p_{\lambda}+\delta
\rho_{\lambda})+3(p_{\lambda}+\rho_{\lambda})\phi'\}\quad ,\nonumber\\
&=&\delta Q^0_{(\lambda)}\nonumber \label{perturbation} \\
\partial_i \delta \nabla_{\mu}T^{\mu i}_{(\lambda)} &=&
\frac{1}{a^2}\{[p'_{\lambda}+\mathcal
{H}(p_{\lambda}+\rho_{\lambda})]\nabla^2B+[(p'_{\lambda}+\rho'_{\lambda})+4\mathcal{H}(p_{\lambda}+\rho_{\lambda})]\theta_{\lambda}\\\nonumber
&&+(p_{\lambda}+\rho_{\lambda})\nabla^2B'+\nabla^2\delta
p_{\lambda}+(p_{\lambda}+\rho_{\lambda})\theta_{\lambda}'+(p_{\lambda}+\rho_{\lambda})\nabla^2\psi\}
= \partial_i \delta Q^i_{(\lambda)} \quad 
\end{eqnarray}
regardless of the anisotropic stress, while $\delta
Q^{i}_{(\lambda)}$ is a new perturbation variable. Considering
that the intrinsic momentum transfer can produce acoustics in the
DM fluid as well as pressure which may resist the attraction of
gravity and hinder the growth of gravity fluctuations during
tightly coupled photon baryon period, in our study we shall
neglect the intrinsic momentum transfer by setting $\delta
Q^{i}_{(\lambda)}=0$. This is a choice of interaction and the
results should not heavily depend on such assumption. Our aim is
to provide examples of both stability and instability in
perturbations.

We construct gauge-invariant quantities by employing
Bardeen's potentials, gauge-invariant density contrast and
velocity
\begin{eqnarray}
\Psi &=& \psi - \frac{1}{a}\left[(-B+\frac{E'}{2})a\right]'\nonumber \\
\Phi &=& \phi - \frac{1}{6}\nabla^2E + \frac{a'}{a}(B -
\frac{E'}{2})\nonumber\\
D_{(\lambda)}&=&\delta_{(\lambda)}-\frac{\rho'_{(\lambda)}}{\rho_{(\lambda)}
\mathcal{H}}(\phi- \frac{1}{6}\nabla^2E)\nonumber\\
V_{(\lambda)} &=& v_{(\lambda)} - \frac{E'}{2}\quad .   
\end{eqnarray}
Choosing a particular gauge, the Longitudinal gauge, by defining
$E=0$, $B=0$, one can find $\Psi=\psi,\Phi = \phi$
\cite{mukhanov}.

For the interacting model we use the perturbed pressure
\cite{maartens}
\begin{equation}
\delta p_d=C_e^2\delta_d\rho_d+(C_e^2-C_a^2)\left[\frac{3\mathcal{H}
(1+w)V_d\rho_d}{k}-a^2Q^0_d\frac{V_d}{k}\right] \label{pertpressure}
\end{equation}
and the interaction as a linear combination of the energy densities
of dark sectors,
\begin{eqnarray}
a^2Q_{m}^0=3\mathcal{H}(\lambda_1\rho_{m}+\lambda_2\rho_d)\nonumber\\
a^2Q_{d}^0=-3\mathcal{H}(\lambda_1\rho_{m}+\lambda_2\rho_d)
\end{eqnarray}
where $\lambda_1$ and $\lambda_2$ are small positive dimensionless
constants. The generality in the choice of the couplings relies on
the generality of the models. In case we had a Lagrangian
formulation the coupling should be fixed. Lacking a Lagrangian we
are free to choose our model. We are going to show that for some
choice of couplings we may achieve stability in curvature
perturbations. Choosing a positive sign for the interaction the
direction of the energy transfer goes from DE to DM, which is
required to alleviate the coincidence problem \cite{Zimdahl2001}
and avoid some unphysical problems such as negative DE density
etc\cite{He08014233, maartens}. In \cite{maartens} it was argued
that it is more natural to assume that the interaction between
dark sectors depends on purely local quantities. Considering the
symmetries of the Friedmann-Robertson-Walker metric, we note that
the interaction can vary only in time, rather than from point to
point. The only time parameter in question is the age. Thus, the
factor ${\cal H}$ appears in our interaction, implying that the
interaction depends on the cosmic time through the global
expansion rate.

By taking Fourier transformation of eq(~\ref{perturbation}), we
get perturbation equations
\begin{eqnarray}
D_m'&=&-kU_m+6\mathcal{H}\Psi(\lambda_1+\lambda_2/r)-3(\lambda_1+
\lambda_2/r)\Phi'+3\mathcal{H}\lambda_2(D_d-D_m)/r\quad ,\\
U_m'&=&-\mathcal{H}U_m+k\Psi-3\mathcal{H}(\lambda_1+\lambda_2/r)U_m\quad ,\\
D_d' & = &
-3\mathcal{H}C_e^2\left\{D_d-\left[3(\lambda_1r+\lambda_2)+
3(1+w)\right]\Phi\right\}-3\mathcal{H}(C_e^2-C_a^2)
\left[\frac{3\mathcal{H}U_d}{k}-a^2Q^0_d\frac{U_d}{(1+w)\rho_dk}\right]
 \nonumber\\
&&-3\mathcal{H}w\left[3(\lambda_1r+\lambda_2)+3(1+w)\right]
\Phi+3\mathcal{H}w
D_d+3w'\Phi+3(\lambda_1r+\lambda_2)\Phi'-kU_d-6\Psi\mathcal{H}(\lambda_
1
r+\lambda_2)\nonumber\\
&&+3\mathcal{H}\lambda_1r(D_d-D_m) \label{density} \\
U_d'& = &-\mathcal{H}(1-3w)U_d+kC_e^2\left\{D_d-3[(\lambda_1
r+\lambda_2)+(1+w)]\Phi\right\} \nonumber\\
&&-(C_e^2-C_a^2)a^2Q^0_d\frac{U_d}{(1+w)\rho_d}+
3(C_e^2-C_a^2)\mathcal{H}U_d+(1+w)k\Psi+3\mathcal{H}
(\lambda_1r+\lambda_2)U_d. \label{velocity}
\end{eqnarray}
where $r=\rho_m/\rho_d$, $U=(1+w)V$. We have taken $\delta
H=0$ by assuming the expansion rate in the interaction to be the global
expansion rate. This is a matter of choice.

In the above, $C_a^2 =w <0$. However, it is not clear what
expression should we have for $C_e^2$. In \cite{maartens} it has
been argued in favor of $C_e^2=1$. This is correct for the scalar
field, but it is not obvious for other cases, especially for a
fluid with a constant equation of state. The most dangerous
possibility, as far as instabilities are concerned, is
$C_e^2=1\not = C_a^2=w<0$ since the first term in the second line
of eq(~\ref{velocity}) can lead to blow up when $w$ close to $-1$.
In spite of such a danger we are considering such a case here.
Assuming $C_e^2=1$, $C_a^2=w$, the above equations can be
rewritten as
\begin{eqnarray}
D_d'&=&(-1+w+\lambda_1r)3\mathcal{H}D_d-9\mathcal{H}^2(1-w)
(1+\frac{\lambda_1r+\lambda_2}{1+w})\frac{U_d}{k}-
kU_d+9\mathcal{H}(1-w)(\lambda_1r+\lambda_2+1+w)\Phi\nonumber\\
    &&+3(\lambda_1r+\lambda_2)\Phi'-6\Psi\mathcal{H}
(\lambda_1r+\lambda_2)-3\mathcal{H}\lambda_1rD_m\quad ,
\label{simdensity}    \\
U_d'&=&2\left \{1+ \frac{3}{1+w}(\lambda_1r+\lambda_2) \right
\}\mathcal{H}U_d+kD_d-3k(\lambda_1r+\lambda_2+1+w)\Phi+ (1+w)k\Psi
\quad .\label{simvelocity}
\end{eqnarray}

By using the gauge-invariant quantity
$\zeta=\phi-\mathcal{H}\delta \tau$ and letting $\zeta_m
=\zeta_d=\zeta$ we get the adiabatic initial condition,
\begin{equation}
\frac{D_m}{1-\lambda_1-\lambda_2/r}=\frac{D_d}{1+w+\lambda_1r+\lambda_2}
\quad .
\end{equation}
The curvature perturbation relates to density contrast by \cite{bb}
\begin{equation}
\Phi=\frac{4\pi Ga^2\sum
\rho_i\{D_g^i-\rho_{i}'U_i/\rho_{i}(1+w_i)k\}}{k^2-4\pi Ga^2\sum
\rho_i'/\mathcal{H}}\quad .
\end{equation}

\begin{figure}
\includegraphics[width=3.2in,height=3.2in]{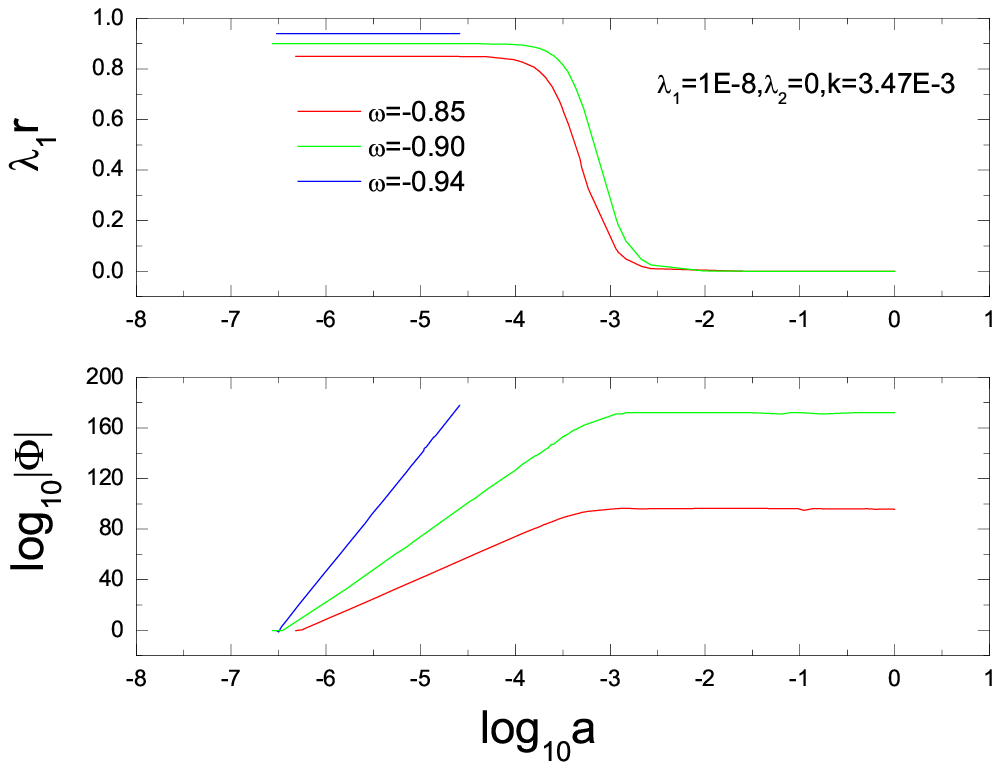}
\includegraphics[width=3.2in,height=3.2in]{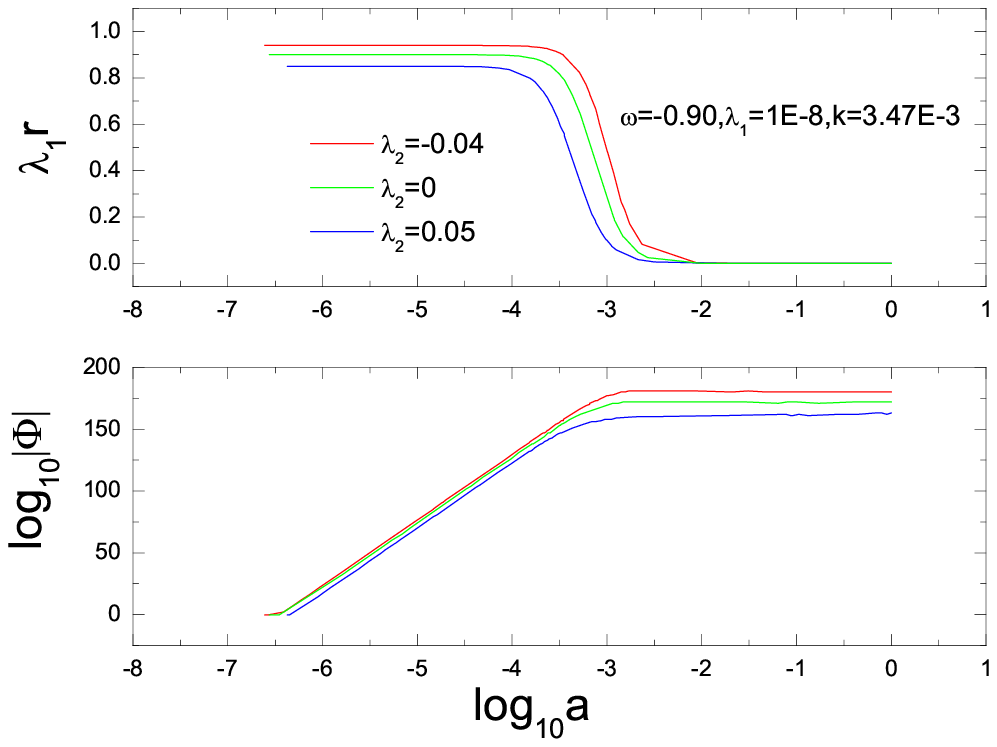}
\caption{The upper two figures show the scaling behavior of
$\lambda_1 r$. The lower two show the behavior of the
perturbation.}\label{figone}
\end{figure}

With the help of these equations we can compute the curvature
perturbation $\Phi$ based on the CMBEASY code. We first consider
the interaction between the dark sectors in proportional to the
energy density of DM ($\lambda_2=0$) and in our calculation we
keep the DE EoS $w\neq -1$. For constant $w>-1$, we show the
numerical results for the  ratio $r=\rho_{m}/\rho_d$ in
Fig.\ref{figone}. We observe that $\lambda_1r$ exhibits a scaling
behavior, which keeps constant both at early and present times of
the universe. This behavior is not changed when we turn on
$\lambda_2$. Analytically, this can be understood by  inserting
the continuity equations
\begin{eqnarray}
\rho'_m+3\mathcal{H}\rho_m &=&
3\mathcal{H}(\lambda_1\rho_m+\lambda_2\rho_d)\nonumber \\
\rho'_d+3\mathcal{H}\rho_d (1+w)&=&
-3\mathcal{H}(\lambda_1\rho_m+\lambda_2\rho_d)
\end{eqnarray}
in
\begin{equation}
r'=\frac{\rho'_m}{\rho_d}-r\frac{\rho'_d}{\rho_d} \, .
\end{equation}
Solving the corresponding quadratic equation, we get
\begin{eqnarray}
(r\lambda_1)_1=-\frac{1}{2}
(w+\lambda_1+\lambda_2)+\frac{1}{2}\sqrt{w^2+2w
\lambda_2+2w\lambda_1+\lambda_2^2-2\lambda_1\lambda_2+\lambda_1^2}
\nonumber \, , \\
(r\lambda_1)_2=-\frac{1}{2}
(w+\lambda_1+\lambda_2)-\frac{1}{2}\sqrt{w^2+2w\lambda_2+
2w\lambda_1+\lambda_2^2-2\lambda_1\lambda_2+\lambda_1^2} \,
.\label{roots}
\end{eqnarray}
This implies
\begin{eqnarray}
(r\lambda_1)_1 &\approx& -(w+\lambda_2)
\nonumber \\
(r\lambda_1)_2 &\approx&
-\frac{\lambda_1\lambda_2}{w+\lambda_2}\sim0\label{simroot}
\end{eqnarray}
 for $\lambda_1 \ll \lambda_2 < - w$. These two roots of
$\lambda_1r$ are constant in the very early time and current time of
the universe, respectively.

The scaling behavior of $\lambda_1r$ influences the curvature
perturbation $\Phi$. Numerically, we see from Fig.\ref{figone}
that when $w>-1$ and $\lambda_1\neq 0$,  $\Phi$ blows up, which
agrees with the result obtained in \cite{maartens}. We find that
this blow-up starts at earlier time when $w$ approaches $-1$ from
above and it happens regardless of the value of $\lambda_2$.

The reason for the blow up is the fact that the expression of $r$
is non perturbative in $\lambda_1$, being proportional to
$\lambda_1^{-1}$ at very times, when we have to consider the
begining of the CMB computation.

However, when we consider the dark sectors' interaction as being
proportional to the energy density of DE and examine the case that
the constant EoS is a little bigger than $-1$: the blow-up
disappears. Stability is also found when we extend our discussion
to the constant EoS $w<-1$.

\begin{figure}
\begin{center}
  \begin{tabular}{cc}
   \includegraphics[width=3.2in,height=3.2in]{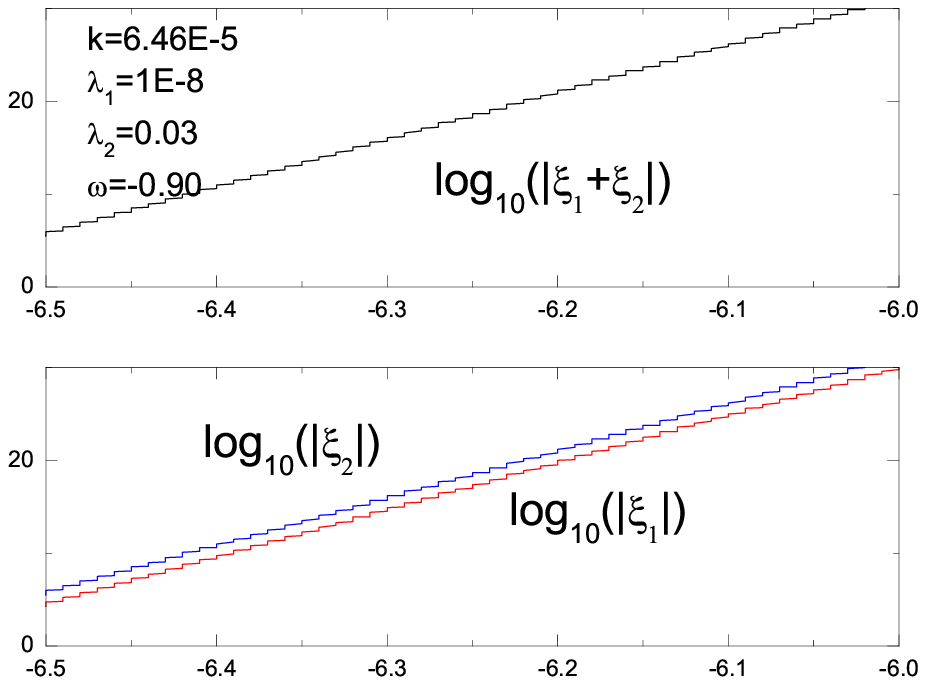}&
   \includegraphics[width=3.2in,height=3.2in]{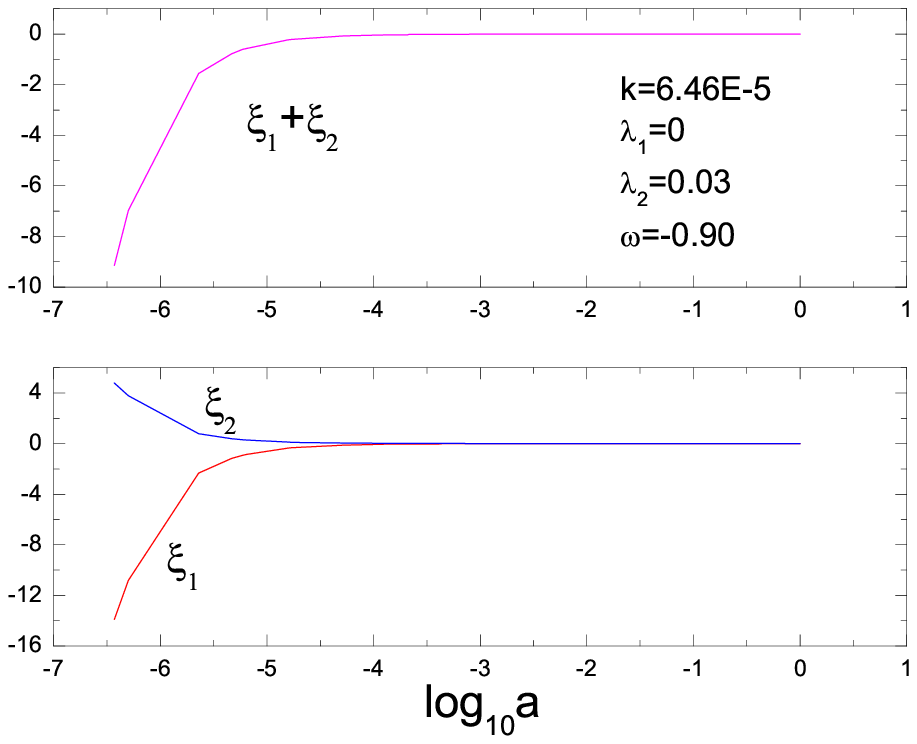}\\
   a&b
  \end{tabular}
  \begin{tabular}{c}
   \includegraphics[width=3.2in,height=3.2in]{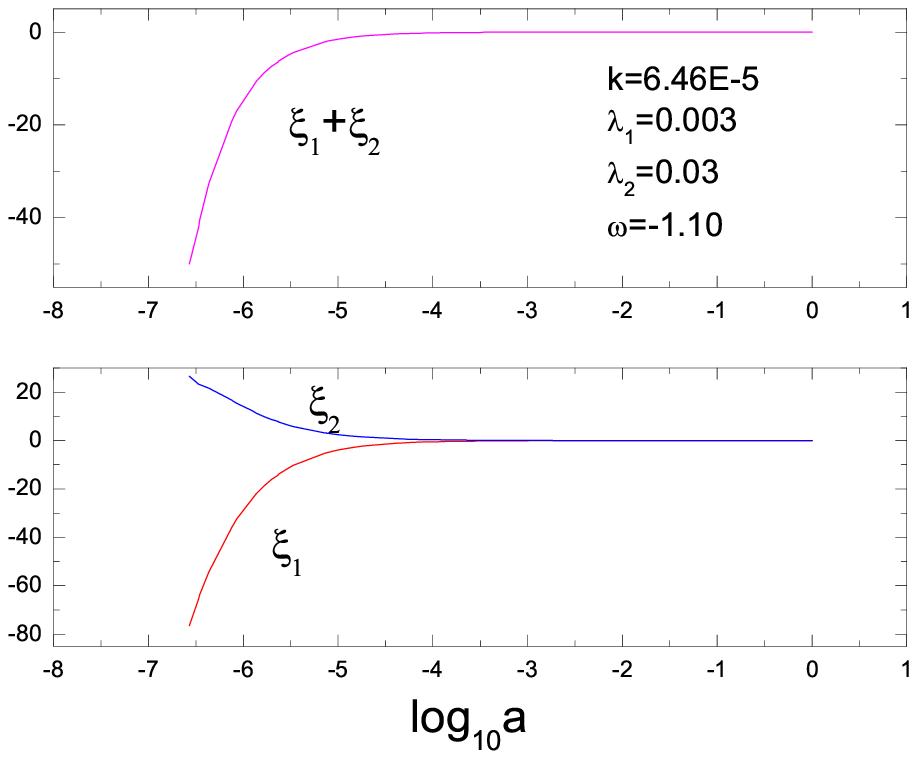}\\
   c
  \end{tabular}
\end{center}
\caption{Comparisons of $\xi_1$ and $\xi_2$.} \label{explain}
\end{figure}

Numerically, we find that the first two terms on the RHS of eqs.
(\ref{simdensity}) and (\ref{simvelocity}) contribute more than
other terms to the divergence. Using $\xi_1$ and $\xi_2$ to
represent the first two terms of eq. (\ref{simdensity}), we can
approximately write
\begin{equation}
D_d'\sim\xi_1+\xi_2\quad ,
\end{equation}
where
\begin{eqnarray}
\xi_1&=&(-1+w+\lambda_1r)3\mathcal{H}D_d\quad , \nonumber \\
\xi_2&=&-9\mathcal{H}^2(1-w)(1+\frac{\lambda_1r+
\lambda_2}{1+w})\frac{U_d}{k}\quad .
\end{eqnarray}
When $\lambda_2=0, \lambda_1\neq 0$ and $-1<w<0$,
$\lambda_1r\approx-w$, we have $\xi_2$ one order larger than
$\xi_1$ and $\xi_1+\xi_2>0$, which causes the vast increase in
$D_d$ as shown in Fig.\ref{explain}a. However, when
$\lambda_1\neq0$ in the case $w<-1$ and $\lambda_1=0$,
$\lambda_2\neq 0$ no matter whether $w<-1$ or $w>-1$, $\xi_2$ and
$\xi_1$ are of the same order as shown in Fig.\ref{explain}b,c and
$\xi_1+\xi_2<0$, which makes $D_d$ to decrease with time.
Therefore, the blow up is avoided.

In order to further explain the reason for the blow-up we provide
an analytical analysis. Keeping the leading terms, we have the
approximate equations
\begin{eqnarray}
D_d'&\approx&(-1+w+\lambda_1r)3\mathcal{H}D_d-9\mathcal{H}^2
(1-w)(1+\frac{\lambda_1r+\lambda_2}{1+w})\frac{U_d}{k} \, , \nonumber\\
U_d'&\approx&2\left[1+\frac{3}{1+w}(\lambda_1r+\lambda_2)\right]
\mathcal{H}U_d+kD_d.\label{approx}
\end{eqnarray}

\begin{figure}
\includegraphics[width=2.8in,height=2.8in]{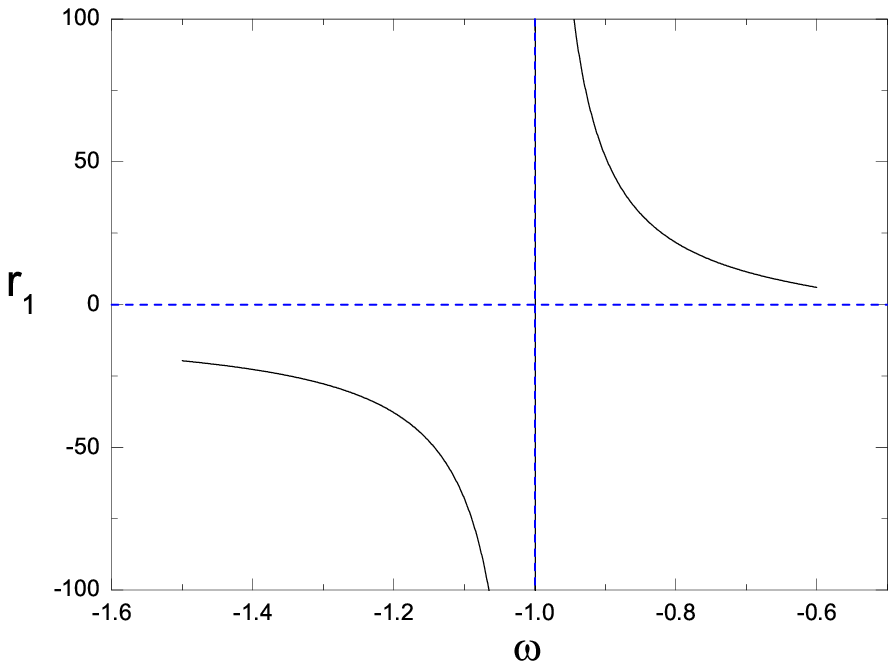}
\includegraphics[width=2.8in,height=2.8in]{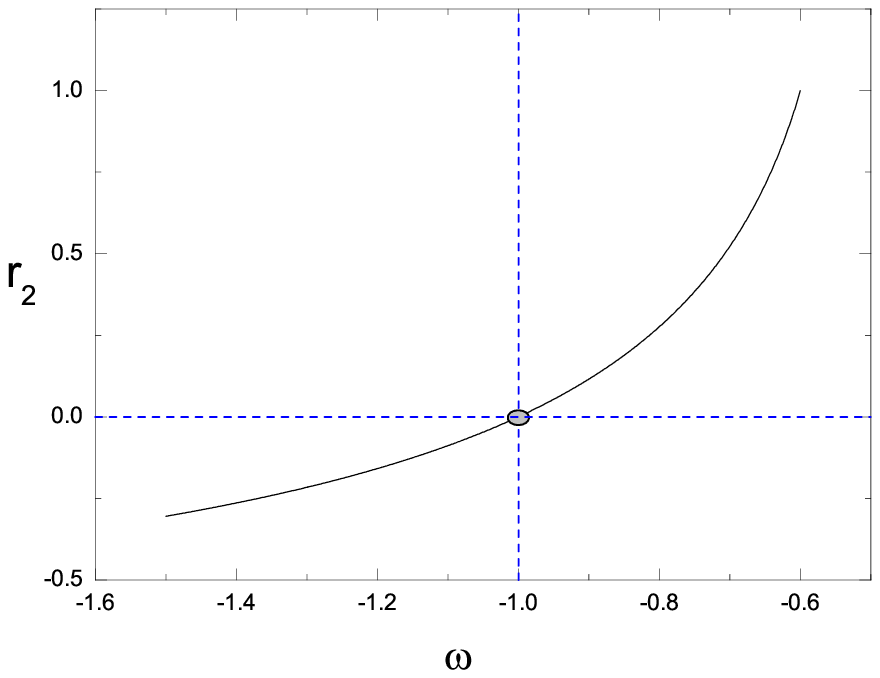}
\caption{Behavior of the indices $r_1$ and $r_2$  in terms of
 $w$. \label{index}}
\end{figure}

Considering the case that the interaction between dark sectors is
proportional to the energy density of DM
($\lambda_1\neq0,\lambda_2=0$) and noting that $\lambda_1r\sim-w$,
we can simplify the above equations to
\begin{eqnarray}
D_d'&\approx&-3\mathcal{H}D_d-9\mathcal{H}^2\frac{1-w}
{1+w}\frac{U_d}{k}\nonumber\\
U_d'&\approx&2\frac{1-2w}{1+w}\mathcal{H}U_d+kD_d\quad .
\end{eqnarray}
 A second order differential equation for $D_d$  is
\begin{equation}
D_d''\approx \left(2\frac{\mathcal{H}'}{\mathcal{H}}-\frac{1+7w}
{1+w}\mathcal{H}\right)D_d'+3(\mathcal{H}'-\mathcal{H}^2)D_d \quad
.\label{secondorder}
\end{equation}
In the radiation dominated period, we have
$\mathcal{H}\sim\frac{1}{\tau},\mathcal{H}'\sim-\frac{1}{\tau^2},
\frac{\mathcal{H}'}{\mathcal{H}}\sim-\frac{1}{\tau}$ and eq.
(\ref{secondorder}) can be approximated as
\begin{equation}
D_d''\approx-3\frac{1+3w}{1+w}\frac{D_d'}{\tau}-\frac{6}{\tau^2}D_d
\quad ,
\end{equation}
whose solution is
\begin{equation}
D_d\approx C_1\tau^{r_1}+C_2\tau^{r_2}\quad \label{growth1},
\end{equation}
where
\begin{eqnarray}
r_1=-\frac{1+4w-\sqrt{-5-4w+10w^2}}{1+w}
\quad ,\nonumber\\
r_2=-\frac{1+4w+\sqrt{-5-4w+10w^2}}{1+w}\quad .\label{growth2}
\end{eqnarray}
The result, eq(\ref{growth2}), has also been given in \cite{maartens}, see
their equations (84) and (85) by setting $\alpha=0$, putting the obvious $+/-$ in
front of the square root and remembering $D_d\sim \psi$. In
\cite{maartens}, $n$ corresponds to $\psi$ while $r_s$ corresponds
to $D$. In fig.\ref{index}, we see that when $-1<w<0$, both $r_1$
and $r_2$ are positive, which means that the perturbation in $D_d$
grows. However, when $w<-1$, both $r_1$ and $r_2$ are negative; this
results in the decay of the perturbation in $D_d$. No divergence
occurs, regardless of the value of $\lambda_1$.

These results tell us that for a constant DE EoS $w>-1$, a
coupling between DE and DM in proportional to $\rho_m$($\lambda_1
\neq 0$) will lead to a violent divergence in the curvature
perturbation. However, this divergence is absent for $ w<-1$.

Considering the case that the interaction between dark sectors is
proportional to the energy density of DE, namely,
$\lambda_1=0,\lambda_2\neq0$, eq(\ref{approx})  reduces to
\begin{eqnarray}
D_d'&\approx&(-1+w)3\mathcal{H}D_d-9\mathcal{H}^2(1-w)
\left(1+\frac{\lambda_2}{1+w}\right)\frac{U_d}{k}\quad , \nonumber\\
U_d'&\approx&2
\left(1+\frac{3\lambda_2}{1+w}\right)\mathcal{H}U_d+kD_d\quad .
\end{eqnarray}
We can rewrite the second order differential equation for $D_d$ in
the form
\begin{equation}
D_d''=\left[\left(-1+3w+\frac{6\lambda_2}{1+w}\right)\mathcal{H}+
2\frac{\mathcal{H}'}{\mathcal{H}}\right]D_d'+3(1-w)\left[
\mathcal{H'}+\mathcal{H}^2\left(-1+\frac{3\lambda_2}{1+w}\right)\right]D_d\quad .
\end{equation}
In the radiation dominated era, the above equation becomes
\begin{equation}
D_d''=\left(-3+3w+\frac{6\lambda_2}{1+w}\right)\frac{D_d'}{\tau}+
3(1-w)\left(-2+\frac{3\lambda_2}{1+w}\right)\frac{D_d}{\tau^2}\quad .
\end{equation}
Introducing the  auxiliary quantities
\begin{eqnarray}
\Gamma&=&3w^2+w+6\lambda_2-2\quad ,\nonumber \\
\Delta&=&9w^4+30w^3+13w^2+(-28+12\lambda_2)w+36\lambda_2^2+
12\lambda_2-20\quad ,
\end{eqnarray}
\begin{figure}
\includegraphics[width=2.8in,height=2.8in]{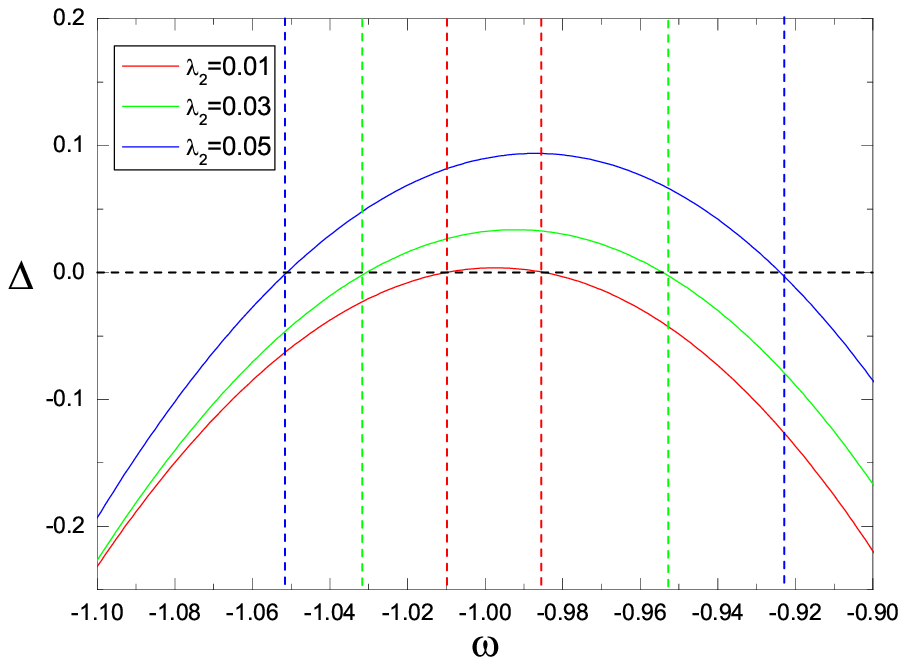}
\includegraphics[width=2.8in,height=2.8in]{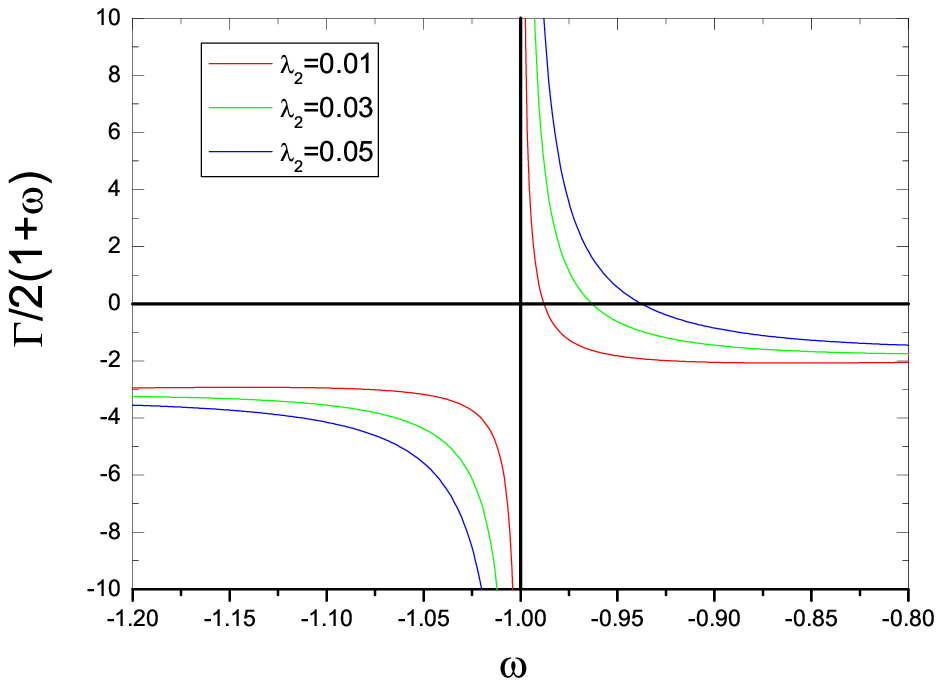}
\caption{Behavior of $\Delta$ and of $\Gamma$. }\label{Delta}
\end{figure}
we find that, when $\Delta>0$,
\begin{equation}
D_d\sim C_1\tau^{r_1}+C_2\tau^{r_2}\quad ,\label{solution}
\end{equation}
where
\begin{eqnarray}
r_1=\frac{1}{2}\frac{\Gamma}{1+w}+\frac{1}{2}
\frac{\sqrt{\Delta}}{1+w}\quad ,\nonumber\\
r_2=\frac{1}{2}\frac{\Gamma}{1+w}-\frac{1}{2}\frac{\sqrt{\Delta}}
{1+w}\quad .
\end{eqnarray}
On the other hand, when $\Delta<0$,
\begin{equation}
D_d\sim C_1\tau^{\frac{1}{2}\frac{\Gamma}{1+w}} cos
\frac{1}{2}\frac{\sqrt{\mid\Delta\mid}}{1+w}ln\tau+C_2
\tau^{\frac{1}{2}\frac{\Gamma}{1+w}} sin
\frac{1}{2}\frac{\sqrt{\mid\Delta\mid}}{1+w}ln\tau \quad
.\label{solution2}
\end{equation}

In fig.~\ref{Delta} we see that $\Delta$ can be positive only in the
vicinity of $w=-1$. When $\lambda_2$ is small, the range for
positive $\Delta$ is small.  $w=-1$ is the central singularity,
since it will lead to the divergence in $r_1$ and cause the blow-up
in the density perturbation eq(\ref{solution}). When $w>-1$ and
$\Delta>0$, the blow-up in the density perturbation can also happen
since $\Gamma/2(1+w)$ is positive as well. But when $w$ grows
further above $-1$, $\Delta$ will become negative and so does
$\Gamma/2(1+w)$, which will lead to the convergent result  of
eq(\ref{solution2}). When $w<-1$, $\Gamma/2(1+w)$ is always
negative, the density perturbation will  decay even when $w$ is
close to $-1$ from below and $\Delta$ is small and positive.

The physical origin of such a behaviour can be traced to eq (\ref{pertpressure}).
When $\lambda_1=0$, the dark energy sound speed depends only on
dark energy parameters, contrary to what happens when $\lambda_1\ne 0$.
In this latter case, the coupling introduces a dependence of the
pressure perturbation on the dark matter energy density. In the latter
case, at early times $\rho_m>>\rho_d$ and the non-adiabatic pressure
perturbation diverges at superhorizon scales, driving the instability.
In our case, the effect is less acute and the system of coupled
differential equations describing the evolution is better behaved.

These results show that when the interaction between dark sectors is
proportional to the energy density of DE($\lambda_2 \neq 0$), the
blow-up in the perturbation will not happen for constant EoS $w<-1$.
For $w>-1$, when the coupling is small, the blow-up can also be
avoided in the observational range of the EoS. However, there is a
possibility for the divergence to happen when the interaction is
large in the observationally allowed $w>-1$ range.

In summary, we have reexamined the cosmological perturbations when
DE and DM  interact with each other. We have specialized the
interaction to be a linear combination of DE and DM energy
densities, namely $\lambda_1\rho_{m}+\lambda_2\rho_d$. We found
that for constant DE EoS $w>-1$ and nonzero $\lambda_1$ the
instability occurs in agreement with the  results of Ref.
\cite{maartens}. However when $w>-1$ and the interaction is just
proportional to the energy density of
DE($\lambda_1=0,\lambda_2\neq0$), the perturbation is stable for
small $\lambda_2$ when $w$ is within observational range. For
phantom DE case with constant $w<-1$, the perturbation is stable
regardless of the value of the coupling. This result was also
evidently shown in \cite{maartens}. It would be interesting to
extend this study to other interaction forms. Moreover, it would
be of great interest to confront the stable DE and DM interaction
model to observations, such as CMB angular power and large scale
structure etc. Works in these directions are in progress.

\acknowledgments{This work has been supported partially by NNSF of
China, Shanghai Science and Technology Commission and Shanghai
Education Commission. The work of E. A. was supported by FAPESP
and CNPQ, Brazil. B. W. would like to acknowledge D. Pavon and F.
A. Barandela for helpful discussions.}

\end{document}